# Zero-dimensional organic exciton-polaritons in tunable coupled Gaussian defect microcavities at room temperature

*Darius Urbonas,[1] Thilo Stöferle,[1,*] Fabio Scafirimuto,[1] Ullrich Scherf[2], and Rainer F. Mahrt[1]*

[1] IBM Research – Zurich, Säumerstrasse 4, 8803 Rüschlikon, Switzerland

[2] Macromolecular Chemistry Group and Institute for Polymer Technology, Bergische Universität Wuppertal, Gauss-Strasse 20, 42119 Wuppertal, Germany

ABSTRACT

We demonstrate strong light-matter interaction at ambient conditions between a ladder-type conjugated polymer and the individual modes of a vertical microcavity with tunable resonance frequencies. Zero-dimensional wavelength-scale confinement for the polaritons is achieved through a sub-micron sized Gaussian defect, resulting in a vacuum Rabi splitting of the polariton branches of $2g$ = 166 meV. By placing a second Gaussian defect nearby, we create a polaritonic molecule with tunnel coupling strength of up to $2J \sim 50$ meV. This platform enables the creation of tailor-made potential landscapes with wavelength-scale dimensions and tunable coupling strengths beyond the thermal energy, opening a route towards room-temperature polariton-based quantum simulators.

Exciton-polaritons are bosonic quasi-particles which arise from strong light matter interaction between photons in optical microcavities with high quality factor and small mode volume and excitons in a variety of embedded active mediums such as semiconductor quantum wells,[1] organic-inorganic perovskites[2] and organic light emitters.[3] Non-equilibrium Bose-Einstein condensation (BEC) of polaritons has been realized in many different quantum well structures,[4] and most recently also with conjugated polymers[5,6] at room temperature. By adding an engineered photonic confinement, zero-dimensional systems have been realized using mesa structures[7] and micropillars.[8] In more complex confining potentials interesting phenomena like one-dimensional condensation,[9] Josephson oscillations[10] and $d$-wave condensation in periodic lattices[11] were observed.

Common to all of these structures is the fixed cavity resonance frequency. Hence, it is not possible to vary the polariton wavefunction between more photonic or more excitonic compositions on the same device. Tunable microcavities in which the distance between the cavity mirrors can be changed in the experiment with nanometer precision and stability are an attractive way to maneuver around this roadblock. In this configuration, transversal confinement can be achieved through hemispherical defects,[12] which has been recently exploited to create tunable cavities with semiconductor quantum wells[13,14] and transition metal dichalcogenide monolayers[15] in the strong light-matter coupling regime at cryogenic temperatures. In order to reach tightest confinement on the wavelength scale, which is important for strong polariton interaction and strongly coupled cavity arrays, nanoscale Gaussian-shaped defects supporting high cavity quality factors $Q > 10^5$ could be used.[16] Moreover, these structures allow precise spatial control of the cavity resonance frequency, enabling the creation of arbitrary potential landscapes and controlled disorder. Additionally, their extreme compactness compared to hemispherical cavities relaxes homogeneity requirements on the active medium, which is critical to realize extend lattices.

Here we demonstrate tunable zero-dimensional organic exciton-polaritons at room temperature in a Gaussian defect microcavity. A spin-coated thin polymer film of methyl-substituted ladder-type poly(p-phenylene) (MeLPPP) acts as active medium on a sputtered distributed Bragg reflector (DBR) for the bottom half of the cavity and a DBR deposited on a Gaussian defect fabricated by means of focused ion beam milling,[17] for the top half (see Methods for more details). The material stack is similar to the one used to achieve room-temperature polariton BEC recently,[5] but the new structure now offers wavelength tunability and wavelength-scale optical potentials. Changing the distance between the two mirrors to bring the cavity frequency into resonance with the polymer exciton, we observe the avoided crossing characteristic for the strong light-matter coupling regime. Furthermore, by creating structures with two overlapping Gaussian defects we create photonic molecules which give rise to effective double well potentials for the polaritons where the tunneling rate, i.e. the coupling strength, is set by the distance between the defects.

RESULTS

A schematic of our tunable microcavity with a single Gaussian defect is shown in Figure 1a. Using a micro-photoluminescence setup we excite the MeLPPP layer off-resonantly with a continuous wave laser outside the DBR stop band at 3.1 eV (see Methods for more details). The emitted light is either detected as real space image with a camera (Figure 1b) or dispersed in a spectrograph equipped with a two-dimensional detector in order to record dispersion diagrams (Figure 1c). We identify discrete modes with spatial intensity profiles similar to Laguerre-Gaussian modes LG$nl$, where $n$ is the radial and $l$ the azimuthal quantum number. The LG00 ground state has a Gaussian spatial profile, the LG01 has a donut shape. Both have an energetically flat dispersion with an intensity maximum for LG00 and a node for LG01 at zero in-plane wavevector (corresponding to angle 0° from normal incidence). Because the excitation beam is larger than the Gaussian defect structure, we additionally observe modes with parabolic dispersion relation originating from the planar Fabry-Pérot microcavity (PC) outside the Gaussian defect. The measurement is done at more than 200 meV red-detuning from the exciton resonance near 2.71 eV in order to reduce the excitonic component and the influence of an absorption tail of uncoupled emitters. Still, the observed line width is broadened and corresponds to a $Q$ factor of ~200, which matches the value that we obtain from three-dimensional finite-difference time-domain (3D FDTD) simulations[18] but is significantly below the $Q$ factor of an empty cavity of 1000 – 2000.

In order to observe the splitting of the cavity mode into lower and upper polariton branches characteristic for the strong light-matter coupling regime, we record transmission spectra with a broadband halogen lamp while tuning the cavity frequency across the polymer exciton energy (Figure 2). As we limit the light source to angles around normal incidence (zero in-plane wavevector), we only observe the PC and the LG00 modes. Both show avoided crossings of the polariton branches with the exciton. However, the upper polariton branch is barely visible due to the small photonic wavefunction component and high background transmission in this wavelength range. From fits with a coupled oscillators model we obtain vacuum Rabi splittings of $2g_{PC}$ = 123 meV and $2g_{LG00}$ = 166 meV, which confirms that the strong zero-dimensional confinement of the photonic mode increases the light-matter interaction significantly. For the fit we neglected the coupling of the cavity mode with the vibronic replicas at 2.9 eV and 3.1 eV, as they are already very close or even beyond the DBR mirror stop band edge. As direct comparison with a purely photonic cavity mode, we performed the same measurement where we replaced the bottom half of the cavity with a DBR mirror without polymer. This shows distinctly different behavior and no avoided crossing (white symbols in Figure 2).

Coupling two zero-dimensional microcavities and creating a photonic molecule is the basic building block towards arbitrary effective potential landscapes for polaritons. In such double well configuration, macroscopic Josephson oscillations were observed with semiconductor micropillars at cryogenic temperatures.[10] The tunnel coupling $J$ through the evanescent fields between the two cavities induces a mode splitting of $2J$ between a "bonding" ($LG00_B$) and an "anti-bonding" mode ($LG00_{AB}$), which in fact thereby cannot be described anymore as Laguerre-Gaussian eigenmodes due to the absence of cylindrical symmetry. Because of the strong mode localization in the Gaussian defect cavities on the order of the wavelength, the coupling distances in our case are also on this length scale, making them almost one order of magnitude smaller than typically encountered in coupled micropillars[19] or coupled tunable hemispherical cavities.[20,21] Figure 3 shows how the mode structure changes with respect to a single Gaussian defect by monitoring the real space emission pattern together with the dispersion relation of the lower polariton branch $LP_{LG00}$.

The coupling strength can be set by changing the distance $d$ between the centers of the Gaussian defects. As shown in Figure 4, it can reach $2J$ ~50 meV at the smallest distance where the anti-bonding $LG00_{AB}$ mode becomes degenerate with the $LG01$ mode in uncoupled Gaussians defects. For comparison, we perform 3D FDTD simulations including the full polymer complex refractive index profile as obtained from ellipsometry. The calculations yield similar coupling strength but the obtained slope of *J* as a function of *d* is slightly steeper than in the experiment. This is most probably caused by the barrier between the Gaussian wells being smoothened and slightly lowered as an artifact of the focused ion beam milling process, and hence the cavities in the fabricated structures experience slightly increased coupling at larger distances.

CONCLUSIONS

In conclusion, we demonstrated zero-dimensional organic exciton-polaritons at room temperature in tunable, wavelength-scale coupled Gaussian defect cavities. Compared to planar Fabry-Pérot cavity modes, the light-matter coupling strength is increased by 35% to 166 meV due to the strong photonic confinement. By realizing a doublet structure, we implement the fundamental building block for strongly coupled, extended polariton lattices. The tunnel coupling strength can be set to exceed the thermal energy. Hence, room temperature quantum simulations with organic polariton condensates become feasible based on the presented experimental platform. Furthermore, higher-order Laguerre-Gaussian modes could offer a new degree of freedom to mimic spin physics with polaritons.

METHODS

**Sample preparation.** For the top half of the cavity, we define a 200 μm diameter mesa structure with 30 μm back-etch on a borosilicate substrate onto which we pattern the Gaussian defects (depth ~ 100 nm, full width at half maximum ~ 1050 nm) by focused ion beam milling. Subsequently, we deposit 6.5 layer pairs of ($Ta_2O_5$/$SiO_2$) DBR with magnetron sputtering. For the bottom half, we deposit 9.5 DBR layer pairs on a borosilicate substrate and a 20 nm $SiO_2$ spacer layer. Subsequently, we spin-coat with 1% solution of MeLPPP (synthesis described in ref. 22) in toluene, resulting in a 35 nm thin film, which is afterwards protected with a 10 nm $SiO_2$ layer.

**Optical characterization.** All measurements are performed in a homebuilt micro-photoluminescence setup that is equipped with XYZ nanopositioning stages on which the cavity halves are mounted. The sample is excited with either a single-mode fiber-coupled continuous wave excitation laser at a wavelength of 405 nm or a multi-mode fiber-coupled halogen lamp for the transmission measurements. The excitation beam is focused through the top mirror with a 100X NA=0.5 apochromatic microscope objective on the microcavity to a spot diameter of 2-3 μm. The light from the cavity was collected through the bottom mirror with a 20X NA=0.5 apochromatic microscope objective and detected either with an imaging spectrograph (for the spectra and the dispersion curves) or a camera (for the real space images), with suitable longpass filters to block the excitation beam but without any polarizer in-between. For the dispersion curves, the spectrograph performs *k*-space imaging, and the entrance slit is set to 50 μm width, extracting a slice along the center axis that connects the Gaussian doublets. For the real space imaging, the light is filtered by tunable bandpass filters to separate out the individual cavity modes.

FIGURES

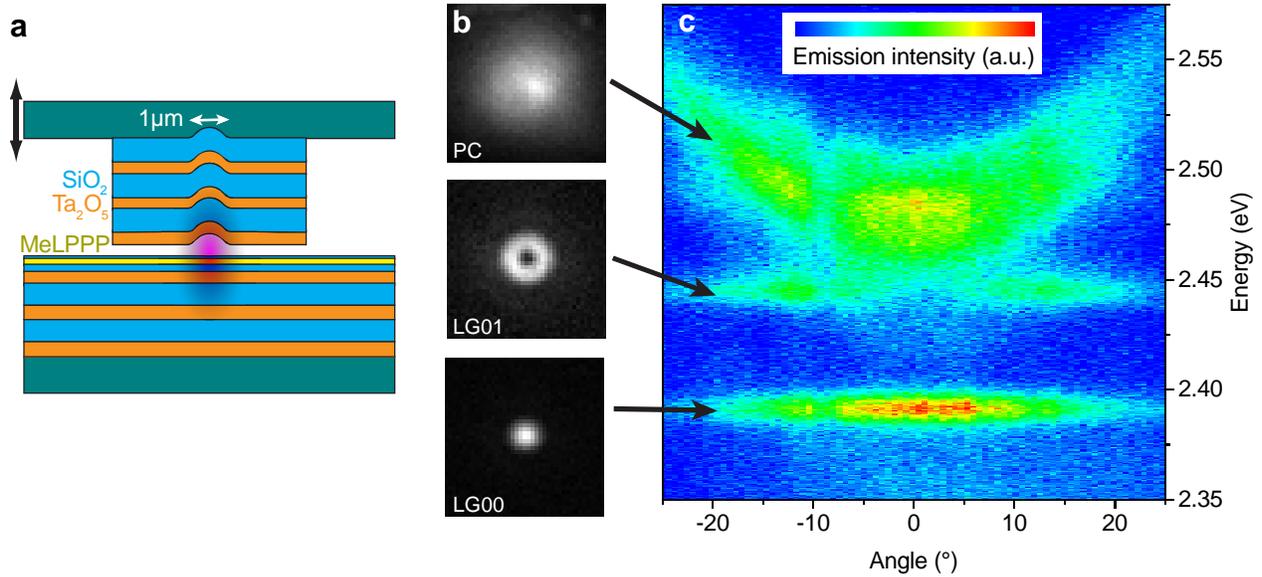

**Figure 1.** (a) Illustration of the tunable Gaussian defect cavity. (b) The resonances of the microcavity with strong lateral confinement have real-space mode patterns of Laguerre-Gaussian modes (LG00 and LG01). The planar Fabry-Pérot cavity resonance (PC) is observed, too. The image gray scales are each normalized to the maximum emission intensity, and the dimensions are 8 μm x 8 μm. The LG00 profile is diffraction limited, exhibiting a perfectly Gaussian profile with 450 nm $1/e^2$ radius. (c) The angular emission is measured directly across the center of the Gaussian structure with an imaging spectrograph in *k*-space imaging configuration and effectively yields the dispersion relations. This shows the flat energy bands and distinct angular distributions for the LG00 and LG01 modes and the parabolic dispersion of the PC mode.

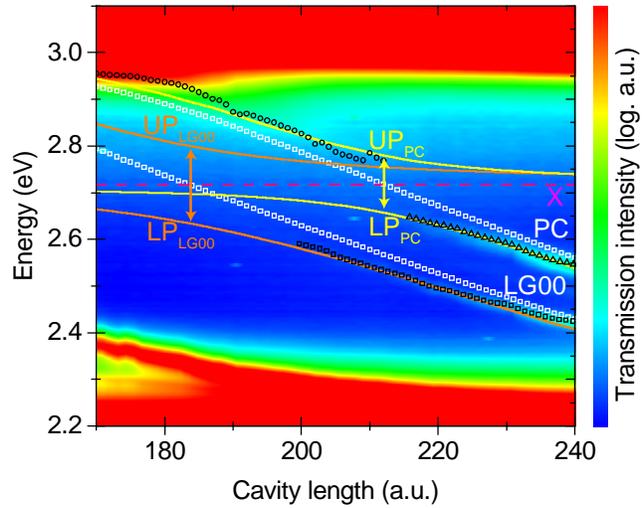

**Figure 2.** The transmission spectrum (color plot) shows the splitting of the localized LG00 mode and the planar cavity (PC) mode into lower (LP) and upper (UP) polariton branches when the cavity length is tuned. For comparison, data from an empty cavity without polymer is shown as white symbols. The solid lines are coupled oscillators fits to the extracted transmission peaks (black symbols) which exhibit the characteristic Rabi splitting when the photonic cavity resonances (LG00/PC) cross the exciton energy (X).

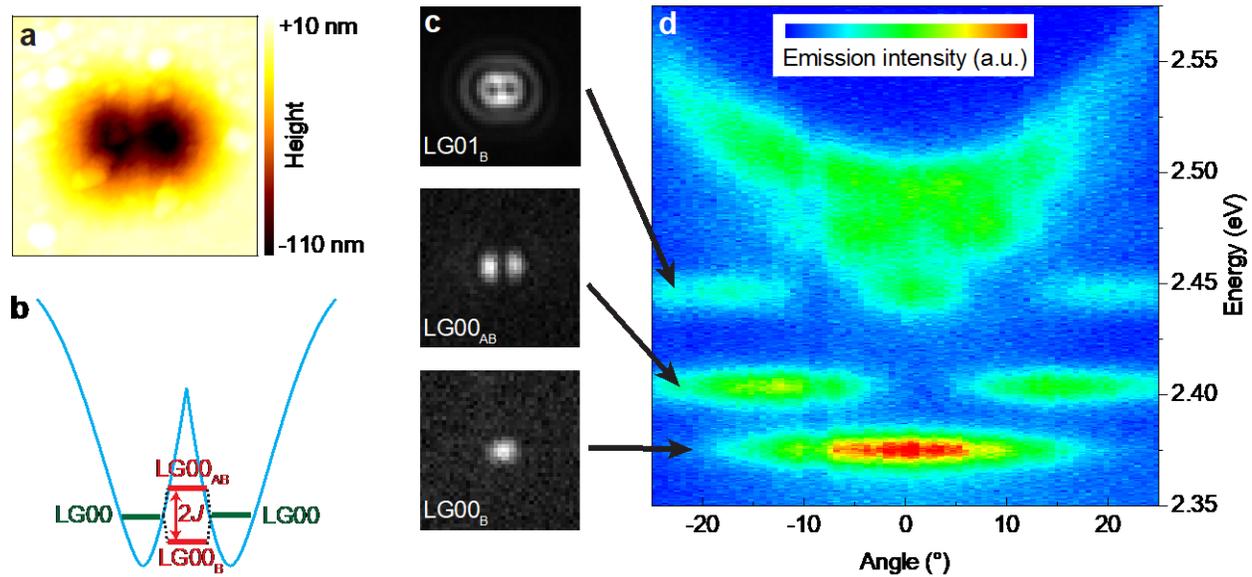

**Figure 3.** (a) Atomic force microscopy image (3 μm x 3 μm) of the fabricated top cavity half with a closely spaced doublet of Gaussians. (b) The tunnel coupling $J$ in the double-well potential splits the LG00 mode into a "bonding" ($LG00_B$) and "anti-bonding" ($LG00_{AB}$) mode. (c) Spatial profiles of the modes (8 μm x 8 μm, intensity normalized). (d) The measured angular emission spectrum shows the characteristic dispersion relations of the different modes.

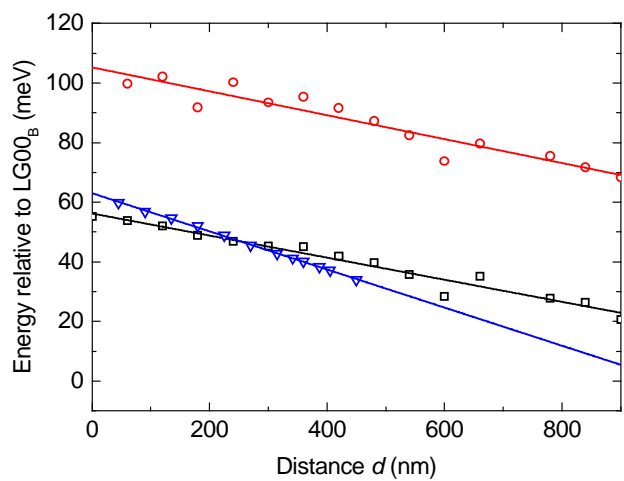

**Figure 4.** The coupling strength is characterized by the splitting between the energy of the LG00$_B$ and LG00$_{AB}$ (black squares) / LG01$_B$ modes (red circles) as a function of distance *d* between the Gaussian defects. Ab-initio 3D FDTD simulations for LG00$_{AB}$ show very similar values (blue triangles) with a slightly different slope which is probably caused by fabrication inaccuracies. The solid lines are linear fits to the data to guide the eye.

AUTHOR INFORMATION

**Corresponding Author**

* tof@zurich.ibm.com (T.S.)

**Author Contributions**

The manuscript was written through contributions of all authors. All authors have given approval to the final version of the manuscript.

**Notes**

The authors declare no competing financial interest.

ACKNOWLEDGMENTS

We are grateful to G. Rainò, P. Seidler, B. J. Offrein, G.-L. Bona and A. Imamoglu for discussions, U. Drechsler, R. Stutz, S. Reidt and M. Sousa for help with the sample preparation and characterization. This work was partly supported by the Swiss State Secretariat for Education,

REFERENCES

(1) Weisbuch, C.; Nishioka, M.; Ishikawa, A.; Arakawa, Y. Observation of the Coupled Exciton-Photon Mode Splitting in a Semiconductor Quantum Microcavity. *Phys. Rev. Lett.* **1992**, 69, 3314–3317.

(2) Fujita, T.; Sato, Y.; Kuitani, T.; Ishihara, T. Tunable Polariton Absorption of Distributed Feedback Microcavities at Room Temperature. *Phys. Rev. B* **1998**, 57, 12428–12434.

(3) Lidzey, D. G.; Bradley, D. D. C.; Skolnick, M. S.; Virgili, T.; Walker, S.; Whittaker, D. M. Strong Exciton-Photon Coupling in an Organic Semiconductor Microcavity. *Nature* **1998**, 395, 53–55.

(4) Byrnes, T.; Kim, N. Y.; Yamamoto Y. Exciton–Polariton Condensates. *Nature Phys.* **2014**, 10, 803–813.

(5) Plumhof, J. D.; Stöferle, T.; Mai, L.; Scherf, U.; Mahrt, R. F. Room-Temperature Bose–Einstein Condensation of Cavity Exciton–Polaritons in a Polymer. *Nature Mater.* **2014**, 13, 247–252.

(6) Daskalakis, K. S.; Maier, S. A.; Murray, R.; Kéna-Cohen, S. Nonlinear Interactions in an Organic Polariton Condensate. *Nature Mater.* **2014**, 13, 271–278.

(7) Idrissi Kaitouni, R.; El Daïf, O.; Baas, A.; Richard, M.; Paraiso, T.; Lugan, P.; Guillet, T.; Morier-Genoud, F.; Ganière, J. D.; Staehli, J. L.; Savona, V.; Deveaud, B. Engineering the Spatial Confinement of Exciton Polaritons in Semiconductors. *Phys. Rev. B* **2006**, 74, 155311.

(8) Bajoni, D.; Senellart, P.; Wertz, E.; Sagnes, I.; Miard, A.; Lemaître, A.; Bloch, J. Polariton Laser Using Single Micropillar GaAs-GaAlAs Semiconductor Cavities. *Phys. Rev. Lett.* **2008**, 100, 047401.

(9) Wertz, E.; Ferrier, L.; Solnyshkov, D. D.; Johne, R.; Sanvitto, D.; Lemaître, A.; Sagnes, I.; Grousson, R.; Kavokin, A. V.; Senellart, P.; Malpuech, G.; Bloch, J. *Nature Phys.* **2010**, 6, 860–864.

(10) Abbarchi, M.; Amo, A.; Sala, V. G.; Solnyshkov, D. D.; Flayac, H.; Ferrier, L.; Sagnes, I.; Galopin, E.; Lemaître, A.; Malpuech, G.; Bloch, J. Macroscopic Quantum Self-Trapping and Josephson Oscillations of Exciton Polaritons. *Nature Phys.* **2013**, 9, 275–279.

(11) Kim, N. Y.; Kusudo, K.; Wu, C.; Masumoto, N.; Löffler, A.; Höfling, S.; Kumada, N.; Worschech, L.; Forchel, A.; Yamamoto, Y. Dynamical d-Wave Condensation of Exciton–Polaritons in a Two-Dimensional Square-Lattice Potential. *Nature Phys.* **2011**, 7, 681–686.

(12) Vijaya Prakash, G.; Besombes, L.; Kelf, T.; Baumberg, J. J.; Bartlett, P. N.; Abdelsalam, M.E. Tunable Resonant Optical Microcavities by Self-Assembled Templating. *Optics Letters* **2004**, 29, 1500–1502.

(13) Dufferwiel, S.; Fras, F.; Trichet, A.; Walker, P. M.; Li, F.; Giriunas, L.; Makhonin, M. N.; Wilson, L. R.; Smith, J. M.; Clarke, E.; Skolnick, M. S.; Krizhanovskii, D. N. Strong Exciton-Photon Coupling in Open Semiconductor Microcavities. *Appl. Phys. Lett.* **2014**, 104, 192107.

(14) Besga, B.; Vaneph, C.; Reichel, J.; Estève, J.; Reinhard, A.; Miguel-Sánchez, J.; Imamoğlu, A.; Volz, T. Polariton Boxes in a Tunable Fiber Cavity. *Phys. Rev. Appl.* **2015**, 3, 014008.

(15) Dufferwiel, S.; Schwarz, S.; Withers, F.; Trichet, A. A. P.; Li, F.; Sich, M.; Del Pozo-Zamudio, O.; Clark, C.; Nalitov, A.; Solnyshkov, D. D.; Malpuech, G.; Novoselov, K. S.; Smith, J. M. ; Skolnick, M. S.; Krizhanovskii, D. N.; Tartakovskii, A. I. Exciton–Polaritons in Van Der Waals Heterostructures Embedded in Tunable Microcavities. *Nature Comm.* **2015**, 6, 8579.

(16) Ding, F.; Stöferle, T.; Mai, L.; Knoll, A.; Mahrt, R. F. Vertical Microcavities with High Q and Strong Lateral Mode Confinement. *Phys. Rev. B* **2013**, 87, 161116(R).

(17) Mai, L.; Ding, F.; Stöferle, T.; Knoll, A.; Offrein, B. J.; Mahrt. R. F. Integrated Vertical Microcavity Using a Nano-Scale Deformation for Strong Lateral Confinement. *Appl. Phys. Lett.* **2013**, 103, 243305.

(18) Oskooi, A. F.; Roundy, D.; Ibanescu, M.; Bermel, P.; Joannopoulosa, J. D.; Johnson, S. G. MEEP: A flexible free-software package for electromagnetic simulations by the FDTD method. *Comp. Phys. Comm.* **2010**, 181, 687–702.

(19) Bayer, M.; Gutbrod, T.; Reithmaier, J. P.; Forchel, A.; Reinecke, T. L.; Knipp, P. A.; Dremin, A. A.; Kulakovskii, V. D. Optical Modes in Photonic Molecules. *Phys. Rev. Lett.* **1998**, 81, 2582–2585.

(20) Flatten, L. C.; Trichet, A. A. P.; Smith, J. M. Spectral Engineering of Coupled Open-Access Microcavities. *Laser & Photonics Reviews* **2015**, 10, 257–263.

(21) Dufferwiel, S.; Li, F.; Trichet, A. A. P.; Giriunas, L.; Walker, P. M.; Farrer, I.; Ritchie, D. A.; Smith, J. M.; Skolnick, M. S.; Krizhanovskii, D. N. Tunable Polaritonic Molecules in an Open Microcavity System. *Appl. Phys. Lett.* **2015**, 107, 201106.

(22) Scherf, U.; Bohnen, A.; Müllen, K. Polyarylenes and Poly(arylenevinylene)s,9 The Oxidized States of a (1,4-Phenylene) Ladder Polymer. *Makromol. Chem.* **1992**, 193, 1127–1133.

Table of Contents Graphic

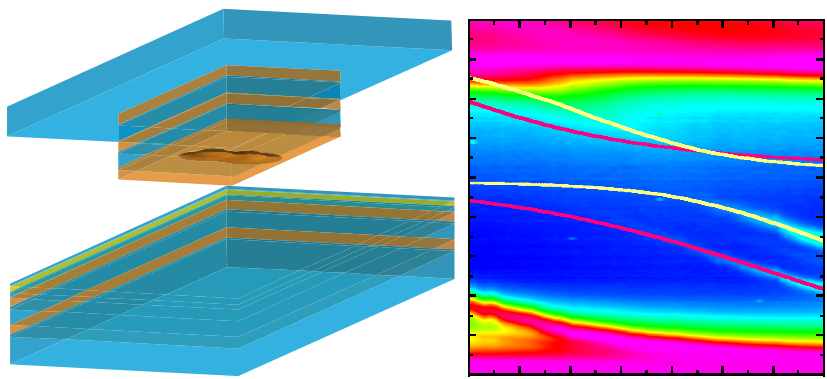